\documentclass[conference]{IEEEtran}
\usepackage[T1]{fontenc}
\usepackage{mathtools}
\usepackage{xurl}
\usepackage{amsmath}
\usepackage{amssymb}
\usepackage{amsthm}
\usepackage{amsfonts}
\usepackage{braket}

\usepackage{upgreek}
\usepackage{dirtytalk}
\DeclarePairedDelimiter\floor{\lfloor}{\rfloor}
\usepackage{optidef}

\ifCLASSINFOpdf
\else
\fi
\usepackage{cite}

\begin{document}
%
\title{Entanglement Rate Optimization in Heterogeneous Quantum Communication Networks}

\author{\IEEEauthorblockN{Mahdi Chehimi and Walid Saad}
\IEEEauthorblockA{Wireless@VT, Bradley Department of Electrical and Computer Engineering, Virginia Tech, Blacksburg, VA USA,\\
Emails: \{mahdic,walids\}@vt.edu}}



%


\maketitle

\begin{abstract}
Quantum communication networks are emerging as a promising technology that could constitute a key building block in future communication networks in the 6G era and beyond. These networks have an inherent feature of parallelism that allows them to boost the capacity and enhance the security of communication systems. Recent advances led to the deployment of small- and large-scale quantum communication networks with real quantum hardware. In quantum networks, \emph{entanglement} is a key resource that allows for data transmission between different nodes. However, to reap the benefits of entanglement and enable efficient quantum communication, the number of generated entangled pairs must be optimized. Indeed, if the entanglement generation rates are not optimized, then some of these valuable resources will be discarded and lost. In this paper, the problem of optimizing the entanglement generation rates and their distribution over a quantum memory is studied. In particular, a quantum network in which users have heterogeneous distances and applications is considered. This problem is posed as a mixed integer nonlinear programming optimization problem whose goal is to efficiently utilize the available quantum memory by distributing the quantum entangled pairs in a way that maximizes the user satisfaction. An interior point optimization method is used to solve the optimization problem and extensive simulations are conducted to evaluate the effectiveness of the proposed system. Simulation results show the key design considerations for efficient quantum networks, and the effect of different network parameters on the network performance.
\end{abstract}


%
\IEEEpeerreviewmaketitle
\vspace{-0.2cm}
\section{Introduction}
\vspace{-0.2cm}
Quantum mechanics has an inherent feature of parallelism that enables performing computations in a superior manner compared to classical computers. Recently, strong quantum computers which verify quantum supremacy were built \cite{quantum_supremacy}, and networks of quantum computing nodes were developed. These advances in quantum technology motivated the research community to focus on designing advanced quantum networks, with different structures, and protocols. Moreover, there has been significant interest in building the required physical hardware like quantum repeaters and switches in order to be able to develop the worldwide quantum internet \cite{quantum_internet2}. 

In particular, quantum communication networks can endow next-generation 6G wireless systems and beyond with greater computational powers, faster communication capabilities, and higher data throughput. Such added features are essential to build intelligent wireless systems that comply with the challenges stemming from the exponential increase in the number of connected devices and the massive volumes of data generated from connected systems \cite{chowdhury2020_6g_quantum_1}. Moreover, quantum cryptography \cite{quantum_cryptography} and advanced quantum communications can provide fool-proof security against cyber attacks in 6G networks \cite{saad_6g}. 

Quantum communication is inherently different from classical communication since it is based on concepts from quantum mechanics that have no classical counterparts. In quantum communication, qubits are sent instead of classical bits. Unlike classical bits which can only take a binary value, qubits can be in any superposition of both \say{0} and \say{1} bits. This superposition nature of quantum communication will potentially enable it to improve the throughput of communication systems in the future \cite{chowdhury2020_6g_quantum_1}. The main quantum resource that enables quantum communication is \emph{quantum entanglement}, or the spooky action at a distance \cite{einstein_spooky}. By utilizing quantum entanglement, information can be communicated between different parties far away from each other. The transmission of qubits to perform quantum communications can be done by either direct transmission, or the \emph{quantum teleportation} protocol. Quantum teleportation starts by sharing entangled pairs of qubits between the two communicating parties, which are stored in quantum memories. When information is to be sent, a Bell state measurement (BSM) operation is performed and information is received by the user \cite{nielsen_book_2002quantum}.  

In terms of real-world implementations, networks of actual quantum hardware have already been implemented at both small- and large-scales, which validated the practicality of quantum communication networks. Examples include the Cambridge quantum network \cite{dynes2019cambridge_quantum_network} and the Tokyo quantum key distribution (QKD) network \cite{sasaki2011field_Tokyo}, among many others. Other real-world implementations included works on satellite-to-ground quantum communication networks which are currently used to perform QKD \cite{liao2018_satellite2}. An example is the Chinese satellite, Micius, dedicated for quantum communications \cite{jianwei2018progress_micius}. This satellite was used to achieve quantum entanglement distribution over a distance of 1200 km \cite{yin2017_satellite3}. However, in order to reap the benefits of quantum networks, several key challenges must be overcome such as optimizing entanglement generation and distribution, scheduling the quantum memory, and developing dynamic routing protocols.
\vspace{-0.1cm}
\subsection{Prior Art}
\vspace{-0.05cm}
A number of prior works~\cite{quantum_network4,Optimal_Remote_Entanglement_Distribution,quantum_queuing_delay,vardoyan2019stochastic_analysis_entanglement_switch,gyongyosi2018multilayer_optimization,humphreys2018deterministic_delivery_remote_entanglement_quantum_network,dahlberg2019link,chakraborty2020entanglement} attempted to investigate some of those challenges. Recent advances in this field led to a major breakthrough by the work in \cite{quantum_network4} that designed a quantum network protocol for end-to-end secure communication between quantum clients on long distances. Moreover, the work in \cite{Optimal_Remote_Entanglement_Distribution} studied the optimal distribution of entanglement remotely. The authors in \cite{quantum_queuing_delay} analyzed the concept of queuing delay in quantum networks by proposing a model for tracking the quantum queuing delay, based on dynamic programming. In particular, the work in \cite{quantum_queuing_delay} developed a memory management policy that results in exponential reduction in the average queuing delay. Also, the work in \cite{vardoyan2019stochastic_analysis_entanglement_switch} considered a network of multiple users in a star topology served by a quantum switch with entangled pairs of qubits. The authors in \cite{vardoyan2019stochastic_analysis_entanglement_switch} analyzed the number of qubits in the quantum memory of a switch and its capacity using Markov chains and queuing theory.

In \cite{gyongyosi2018multilayer_optimization}, the authors studied a multilayer optimization problem considering both the quantum and classical layers of the quantum internet. For the quantum layer, the total usage of the quantum memories is optimized and the entanglement throughput is maximized, while using the minimum number of entangled links in the network. The authors in \cite{humphreys2018deterministic_delivery_remote_entanglement_quantum_network} studied large-scale quantum networks and analyzed the deterministic delivery of entanglement remotely. They demonstrated this generation using diamond spin qubits where entanglement generation rates are greater than decoherence rates. The work in \cite{dahlberg2019link} proposes a link layer protocol for quantum networks where a stack network is designed for entanglement distribution in multiple scenarios. On top of these works, the authors in \cite{chakraborty2020entanglement} studied the problem of optimizing the rate of distributing entangled pairs of qubits achieved in a quantum network with repeaters. The problem is formulated as a linear programming problem to maximize the rates while satisfying quality requirements.

However, most of these prior works \cite{quantum_network4,Optimal_Remote_Entanglement_Distribution,quantum_queuing_delay,vardoyan2019stochastic_analysis_entanglement_switch,gyongyosi2018multilayer_optimization,humphreys2018deterministic_delivery_remote_entanglement_quantum_network,dahlberg2019link,chakraborty2020entanglement} focus on simple quantum networks with homogeneous users whose requirements of entanglement are the same and where the same entanglement generation rates are achieved for all users. Moreover, works that consider different entanglement rates for different users, such as \cite{dahlberg2019link} and \cite{chakraborty2020entanglement}, either consider them from a link layer protocol point of view without studying an optimization framework \cite{dahlberg2019link}, or try to maximize the rates over network paths while assuming a perfect quantum memory \cite{chakraborty2020entanglement}. For instance, none of these works considers optimizing the rates of entanglement generation for the different users in the network in order to optimally utilize the available quantum memory capacity. This is an important problem for 6G (and beyond) and the quantum Internet that must be addressed since entanglement is a fundamental resource in quantum communications. 
\vspace{-0.2cm}
\subsection{Contributions}
\vspace{-0.1cm}
The main contribution of this paper is a novel framework for optimizing the entanglement generation rates of heterogeneous quantum users in a way that efficiently utilizes the available quantum memory and achieves fairness between users. The contributions of this work can be summarized as follows:
\begin{itemize}
    \item We consider a heterogeneous quantum network with a single central quantum node and multiple users at different distances and with various applications and entanglement rates requirements that will be optimized. This is in stark contrast to prior art \cite{quantum_network4,Optimal_Remote_Entanglement_Distribution,quantum_queuing_delay,vardoyan2019stochastic_analysis_entanglement_switch,gyongyosi2018multilayer_optimization,humphreys2018deterministic_delivery_remote_entanglement_quantum_network} which do not optimize the entanglement generation rates with the memory management. 
    
    \item We formulate and solve a mixed integer nonlinear programming (MINLP) optimization problem whose goal is to maximize the weighted sum of successfully generated entanglement pairs of qubits for all users in the network such that minimum required service is achieved for each user, fairness between users is guaranteed, and quantum memory capacity is efficiently utilized.
    
    \item Simulation results validate the importance of the proposed system model in designing efficient quantum networks where the available resources are optimally utilized while achieving fairness among the different users. The results show the limiting factors in designing efficient quantum networks and the effect of network parameters on the performance.
\end{itemize}

The rest of this paper is organized as follows. Section \ref{sec_system_model} describes the proposed system model. Next, the formulated optimization problem is presented in Section \ref{sec_optimization_problem}. In Section \ref{sec_experiments}, we conduct extensive simulations and experiments and analyse the key results. Finally, conclusions are drawn in Section \ref{sec_conclusion}.
\vspace{-0.1cm}
\section{System Model}
\label{sec_system_model}
\begin{figure}[t]
\begin{center}
\centerline{\includegraphics[width=\columnwidth]{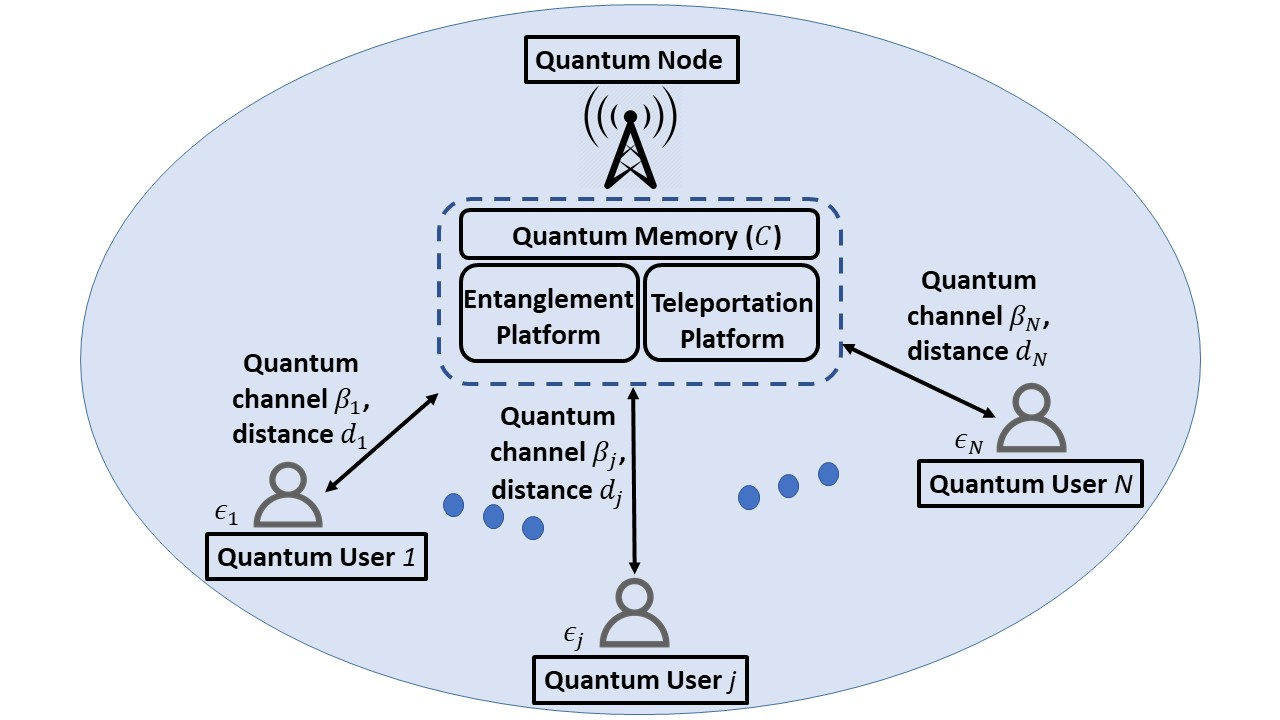}}
\vskip -0.1in
\caption{System Model}
\label{fig_system_model}
\end{center}
\vskip -0.5in
\end{figure}

\vspace{-0.1cm}
Consider the downlink of a quantum network that has a single quantum node serving a set $\mathcal{N}$ of $N$ users located at different distances from the quantum node. The set of users is \emph{heterogeneous} in the sense that they have diverse applications and perform different tasks. Each user requires a specific minimum entanglement generation rate and fidelity level that must be guaranteed in order to satisfy application requirements. The studied system model is illustrated in Figure \ref{fig_system_model}. We do not consider quantum repeaters or switches in our system since we are assuming a small-scale quantum communication network.
The quantum node has an entanglement generation platform that consists of multiple quantum optical hardware dedicated to the generation of photonic entangled Bell pairs of qubits. There are four Bell pairs, and they are maximally entangled pairs of qubits that represent a superposition of the $\ket{0}$ and $\ket{1}$ quantum states. They take one of the following forms: 
\vspace{-0.05in}
\begin{equation}
    \ket{\phi_\pm} = \frac{1}{\sqrt{2}} (\ket{00} \pm \ket{11}),
\end{equation}
\begin{equation}
    \ket{\psi_\pm} = \frac{1}{\sqrt{2}} (\ket{01} \pm \ket{10}).
\end{equation}

Moreover, the quantum node has a quantum memory \cite{lvovsky2009optical_quantum_memory} with capacity to hold $C$ qubits simultaneously. Finally, a quantum teleportation platform composed of Bell state measurement (BSM) devices is included in the quantum node. In the teleportation platform, the quantum teleportation \cite{bouwmeester1997experimental_quantum_teleportation} protocol is performed in order to send information from the quantum node to users. 

During every time duration $\tau$, $N$ devices in the entanglement generation platform generate entangled photon pairs at different rates $r_j$ (pairs/sec), $j \in \mathcal{N}$ dedicated for the $N$ different users. The entangled pairs for different users are generated simultaneously, while, for each user, entangled pairs are generated sequentially using a dedicated hardware. During the time duration $\tau$, a total of $r_j\tau$ entangled pairs of photons is generated for user $j \in \mathcal{N}$. For each entangled pair of photons, one photon is sent over the quantum channel \cite{gyongyosi2018survey_quantum_channel} (e.g. optical fiber link) to its dedicated user. Simultaneously, the other photon is kept in the quantum node and moved to the quantum memory. This photon is either stored in the quantum memory, or used in the teleportation process to send information to its user. The teleportation process destroys the qubit (photon in our case) that is used, since quantum measurement causes the quantum states to collapse \cite{nielsen_book_2002quantum}.

Every photon from an entangled pair that is sent to a user $j \in \mathcal{N}$ at a distance $d_j$ goes over a quantum channel with attenuation coefficient $\beta_j$. The probability of successfully sending such a single photon and identifying it by the user after measurement is given by \cite{vardoyan2019stochastic_analysis_entanglement_switch}:
\begin{equation}
\vspace{-0.1cm}
    P_{s_1,j}(r_j) = e^{-\beta_j d_j}.
    \vspace{-0.1cm}
\end{equation} 
Since the $r_j\tau$ photons generated for user $j \in \mathcal{N}$ are sent one by one independently from one another, we can assume, without loss of generality, that all photons sent to user $j$ undergo the same channel attenuation. Thus, the number of successfully sent photons from the $r_j\tau$ generated pairs of photons to each user $j \in \mathcal{N}$ is
\begin{equation}
\vspace{-0.1cm}
    E = r_j\tau e^{-\beta_j d_j}.
    \vspace{-0.1cm}
\end{equation}
Meanwhile, the other photons which are sent to the quantum memory experience another source of losses. These losses come from the decoherence of qubits and the interactions with the quantum memory \cite{decoherence1,decoherence2}. This is represented by the entangled-state decoherence rate $r_\textrm{dec}$. The duration of time, $\tau$, spent in generating entangled pairs is considered as a parametrization of the decoherence rate where $\tau = \alpha/r_\textrm{dec}$ for some constant $\alpha$. The length of the duration $\tau$ determines the efficiency of the quantum link, since the entanglement generation rate must be larger than the entangled-state decoherence rate in order to preserve the generated pairs. This forms a lower bound on $r_j$, \ $\forall j \in \mathcal{N}$, all of which must be greater than $r_\textrm{dec}$ \cite{humphreys2018deterministic_delivery_remote_entanglement_quantum_network}.

The probability of success in preserving the generated entangled pairs on the quantum node from decoherence when time duration $\tau$ is spent in the generation process is defined as \cite{humphreys2018deterministic_delivery_remote_entanglement_quantum_network}:
\begin{equation}
    P_{s_2,j}(r_j) = 1 - e^{-r_j\tau}, \hspace{0.5cm} \forall j \in \mathcal{N}
\end{equation}

Thus, the total number of successfully generated entangled pairs for user $j \in \mathcal{N}$ is:
\begin{equation}
    S_j = r_j\tau e^{-\beta_j d_j}(1 - e^{-r_j\tau}),
\end{equation}
and this number must be optimized to satisfy each user $j$.

\vspace{-0.2cm}
\section{Entanglement Generation Rates Optimization}
\label{sec_optimization_problem}

Given this model, in order to satisfy the requirements of the different users during a specific time duration, and to save quantum resources, one should not consider the entanglement generation rates as fixed values that are not optimized as done in \cite{humphreys2018deterministic_delivery_remote_entanglement_quantum_network}. In particular there is a need for the quantum node to optimize the rates of entanglement generation dedicated to every user in the network. The goal is to find an optimal distribution of the generated entangled pairs for the different users in the available quantum memory, such that the transmission capacity offered by the heterogeneous network is efficiently utilized and user satisfaction is guaranteed. The quantum node must also achieve $\emph{fairness}$ among users each of which has a minimum integer entanglement generation rate $\epsilon_{\textrm{min},j}$, $\forall j \in \mathcal{N}$. This value represents the minimum entanglement generation rate for each $r_j$ in order to satisfy the application needs for user $j \in \mathcal{N}$. Moreover, due to limitations in the physical hardware, there is an upper limit on the achievable entanglement generation rates that the hardware cannot exceed, which we represent by $\epsilon_{\textrm{max},j}$, $\forall j \in \mathcal{N}$. Let $\boldsymbol{r} = [r_1, r_2,...,r_N]$ be the vector of entanglement generation rates for the $N$ users in the network, which is our optimization variable. The entanglement generation optimization problem can then be formulated as follows:
\begin{subequations}\label{optimization_problem}
\begin{gather}\tag{\ref{optimization_problem}}
\max_{\boldsymbol{r}} \quad \sum_{j=1}^{N}{w_jP_{s_1,j}(r_j)P_{s_2,j}(r_j)r_j\tau},\\    
\begin{align}
\textrm{s.t.} \quad & \sum_{j=1}^{N}{\floor*{r_j\tau P_{s_2,j}(r_j)} = C},\\
  &\epsilon_{\textrm{min},j} \leq r_j \leq \epsilon_{\textrm{max},j},   \hspace{0.7cm} \forall j \in \mathcal{N},
\end{align}
\end{gather}
\end{subequations}
where $w_j$ represents a weighting vector for the entanglement generation rate of user $j$. This weight represents the importance of the task performed by a given user $j$, and it can be controlled to guarantee user service even when high channel losses exist. By efficient weighting of the different users, fairness between users can be achieved and the quantum node will no longer only serve the users closest to it. Constraint ($7\text{a}$) guarantees that throughout the whole entanglement generation process, the quantum memory capacity is never exceeded. This is crucial since exceeding the memory capacity will result in discarding some generated entangled pairs, which represents lost resources. The constraint includes a floor function because the number of generated photon pairs must be an integer. 

The formulated optimization problem is a mixed integer nonlinear programming (MINLP) optimization problem, since variable $\boldsymbol{r}$ must be an integer because we cannot have a partial photon. Generally, MINLP problems are not convex because of their discrete nature. However, an MINLP problem is often considered to be convex if it corresponds to a convex NLP problem after continuous relaxation \cite{MINLP}. For our problem, the Hessian of the objective function consists only of diagonal elements that take the form $(2\tau^2 - r_j\tau^3)e^{-(r_j\tau + \beta_jd_j)}$, $j \in \mathcal{N}$. Thus, the objective function is convex in the region $r_j \leq \frac{2}{\tau}$, and concave in the region $r_j \geq \frac{2}{\tau}$, $\forall j \in \mathcal{N}$. Since $\tau = \alpha/r_\textrm{dec}$, and we must always have $r_j > r_{\textrm{dec}}$, $\forall j \in \mathcal{N}$. Then, whenever we have $\alpha > 2$, we will always be operating in the region where the objective function is concave and has a global maximum. Thus, we will strict our time duration $\tau$ to always maintain $\alpha > 2$.  

MINLP problems can be solved by different algorithms, one of the most famous methods is the branch and bound method \cite{lawler1966branch_and_bound} which, generally, relaxes the integer restriction of the MINLP problem, and solves the resulting (convex) continuous NLP problem. If all resulting values of the originally integer variable take integer values, then the obtained solution is optimal for the MINLP problem \cite{MINLP}. The continuous NLP problem can be solved by different regular optimization algorithms. Here, in order to find a solution for (\ref{optimization_problem}), we use the advanced process monitor (APMonitor) software \cite{Hedengren2014_APMonitor}. This is a specialized optimization software that offers a variety of solvers for mixed-integer problems like the IPOPT, BPOPT, and APOPT solvers. We solve the NLP problem using the interior-point optimization method (IPOPT) solver which is one of the strongest available solvers. By solving the proposed optimization problem, we were able to efficiently optimize the entanglement generation process in the quantum network in order to satisfy all users and to effectively utilize the available quantum memory.  
\vspace{-0.1cm}
\section{Simulation Results and Analysis}
\label{sec_experiments}
\vspace{-0.05cm}
In this section, extensive simulations and experiments are run in order to solve the proposed problem and to verify the effectiveness of the adopted approach.

We begin with defining the ranges of the parameters in the optimization problem and make some design choices to simplify the analysis. First of all, the entanglement generation rates are in the range of giga ebit per second, where an ebit is one unit of bipartite entanglement that corresponds to one of the maximally entangled qubits in the Bell state \cite{vardoyan2019stochastic_analysis_entanglement_switch}. Moreover, each user $j$ is assumed to be connected to the quantum node by a single-mode optical fiber communication link with loss coefficient $\beta_j = 0.2$ dB/km \cite{vardoyan2019stochastic_analysis_entanglement_switch}. For simplicity, we assume the same loss coefficient for all users in the network i.e. $\beta_1 = \beta_2 = ... = \beta_N = 0.2$ dB/km. The users are unequally distant from the quantum node and the distances are in the range of few kilometers. For the quantum memory capacity, typical values normally range between 15 and 50 qubits \cite{quantum_queuing_delay}. In our experiments, we assume a quantum memory capacity of 35 qubits, unless stated otherwise.  

The rate of decoherence is assumed to be 1 giga ebit/sec. For the time duration $\tau$, the parameter $\alpha$ considered is assumed to be 3, unless stated otherwise. Thus, the considered time duration is $\tau = 3$ ns. We assume that the minimum entanglement generation rate is $\epsilon_{\text{min},j} = 1.2\times10^9$ ebit/sec, and the upper limit is $\epsilon_{\text{max},j} = 10\times10^19$ ebit/sec, $\forall j \in \mathcal{N}$. However, some users require a larger minimum value $\epsilon_{\text{min},j}$ necessary for their application, which will be specified during the experiments. Next, we present the conducted experiments and analyze their results.

\subsection{Impact of the Lower Bound of Entanglement Rates}
\begin{figure}[t!]
\begin{center}
\centerline{\includegraphics[scale=0.23]{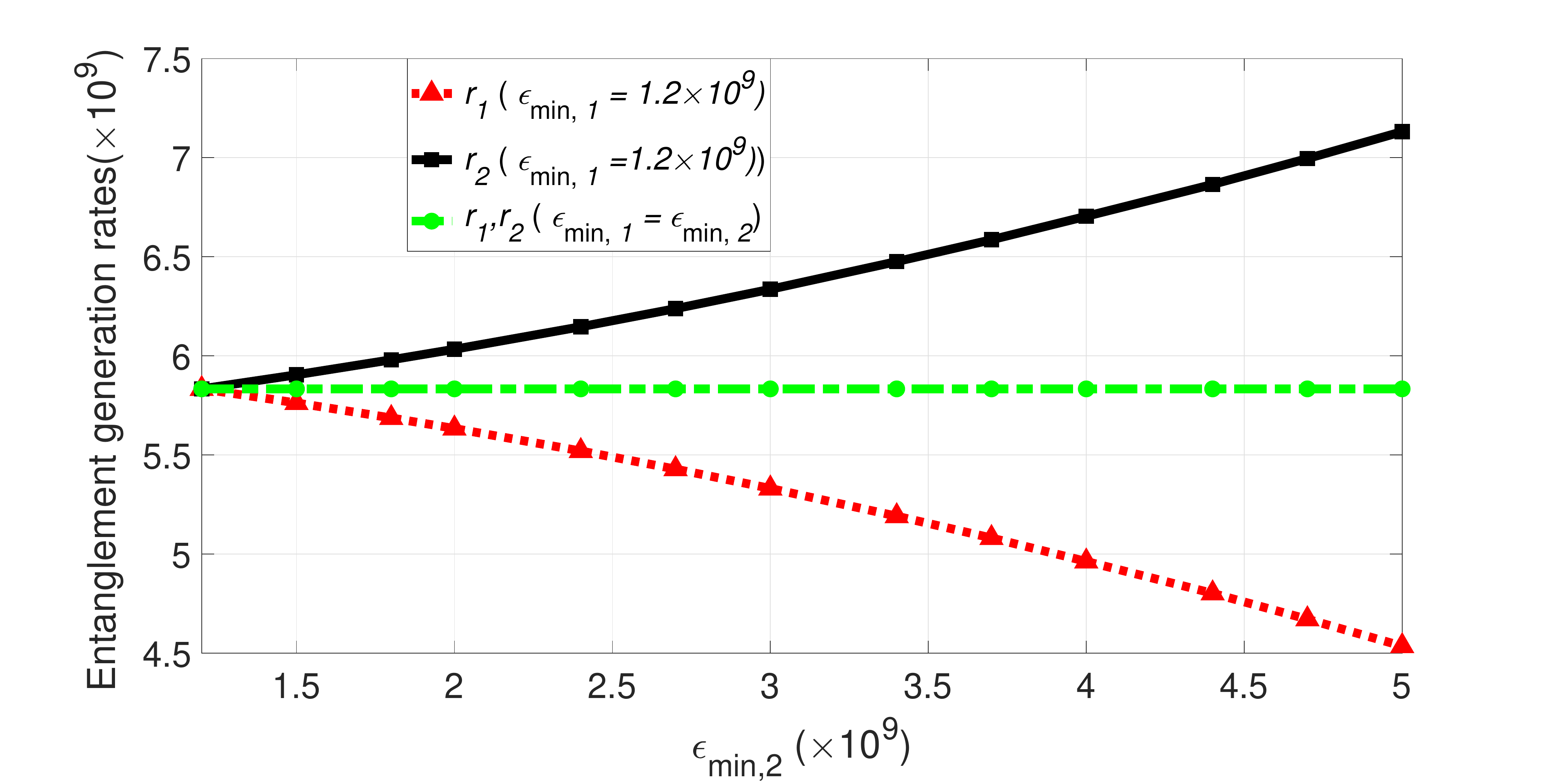}}
\vskip -0.15in
\caption{$\epsilon_{\text{min},2}$ vs $r_1$ and $r_2$ for different values of $\epsilon_{\text{min},1}$, all in (ebit/sec).}
\label{fig1}
\end{center}
\vskip -0.45in
\end{figure}

We begin our experiments by considering a network of two users, with equal distance $d_1 = d_2 = 2$ km from the quantum node and with equal weights: $w_1 = w_2 = 1$. We fix $\epsilon_{\text{min},1} = 1.2\times10^9$ ebit/sec, and vary $\epsilon_{\text{min},2}$ between $1.2\times10^9$ and $5\times10^9$ ebit/sec. We notice that the objective function always evaluates to the same value  which is 4.521, regardless of the values of $\epsilon_{\text{min},1}$ and $\epsilon_{\text{min},2}$. However, the entanglement generation rates for both users vary as $\epsilon_{\text{min},2}$ is varied, which is shown in Figure \ref{fig1}. 
It is clear that the user that requires a higher minimum value of entanglement generation is assigned a larger rate, while the other user's rate gets smaller. This satisfies the fairness criterion since both users are getting entanglement rates larger than their required minimum. Since one rate increases and the other decreases in a similar manner, the objective function maintains its value. Moreover, if both $\epsilon_{\text{min},1}$ and $\epsilon_{\text{min},2}$ exceed $5.8333$, then there is no feasible solution for this case under the available quantum memory capacity of $C=35$ and the chosen time duration $\tau = 3$ ns.

\subsection{Impact of The Time Duration $\tau$}
\begin{figure}[t!]
\vskip -0.1in
\begin{center}
\centerline{\includegraphics[scale=0.25]{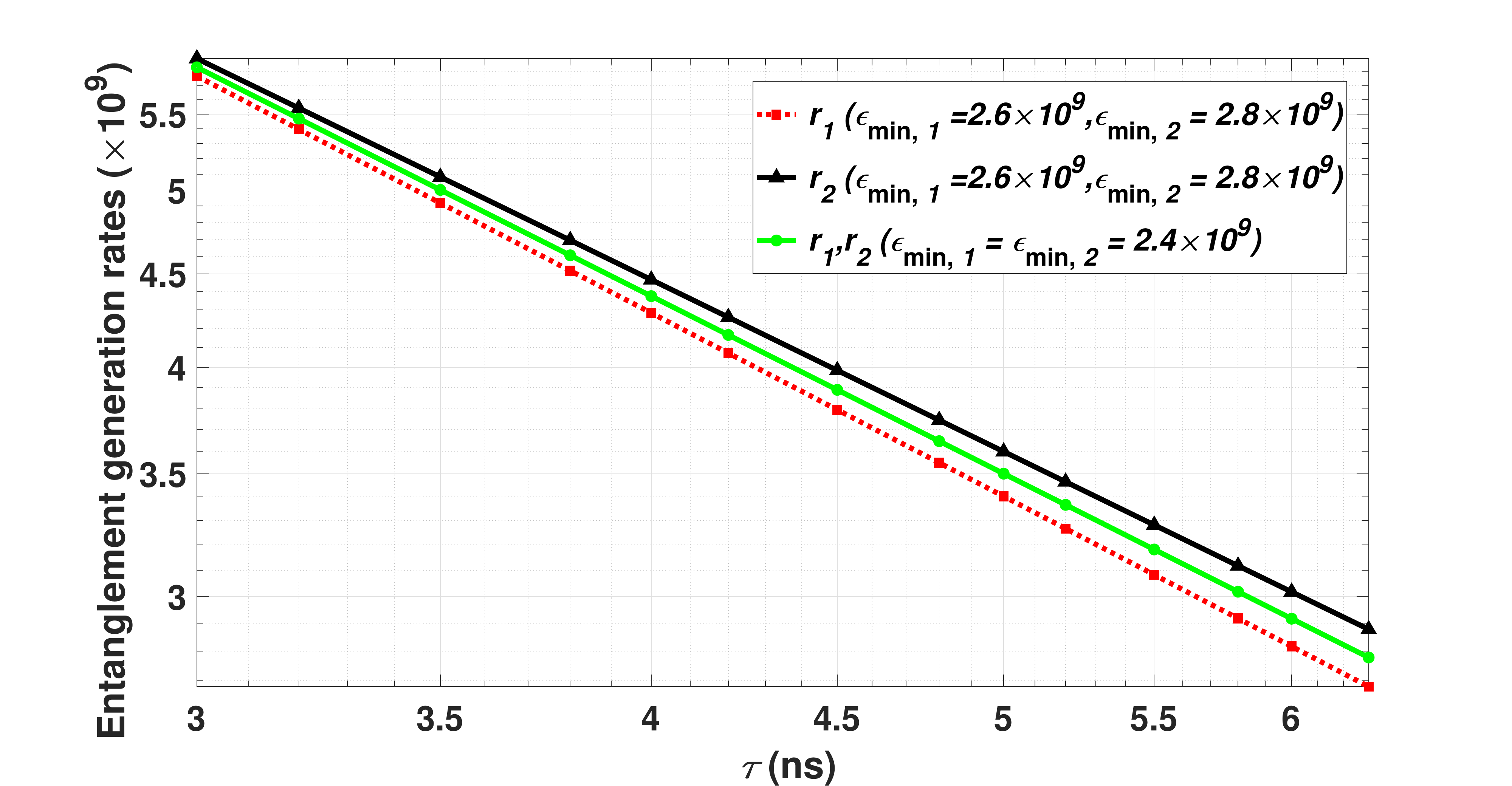}}\vskip -0.15in
\caption{$\tau$ (ns) vs $r_1$ and $r_2$ in ebit/sec for different values of $\epsilon_{\text{min},1}$ and $\epsilon_{\text{min},2}$ in (ebit/sec).}
\label{fig2}
\end{center}
\vskip -0.35in
\end{figure}

In Figure \ref{fig2}, we study the effect of varying the time duration $\tau$ spent in the entanglement generation process. Again, we consider a network of two users with $d_1 = d_2 = 2$ km and $w_1 = w_2 = 1$.
We considered two scenarios with different values of $\epsilon_{\text{min},1}$ and $\epsilon_{\text{min},2}$. In the first scenario, we set $\epsilon_{\text{min},1} = 2.6\times10^9$ and $\epsilon_{\text{min},2} = 2.8\times10^9$. From the figure, we can see that $r_1$ and $r_2$ decrease as $\tau$ increases. This is due to the fact that increasing $\tau$ means generating more entangled pairs, which would result in exceeding the quantum memory capacity if the entanglement generation rates were not decreased. Also, since user 2 has a larger minimum value, it is always assigned a rate that is greater than the rate of user 1. Also, in this scenario, for $\tau > 6.4814$ ns, there is no feasible solution that satisfies the constraints on the quantum memory capacity $C=35$. In the second scenario, we set $\epsilon_{\text{min},1} = \epsilon_{\text{min},2} = 2.4\times{10^9}$ ebit/sec. As expected, $r_1$ and $r_2$ decrease as $\tau$ increases, but now both users are getting the same rate. Here, for $\tau > 7.2916$ ns, there is no feasible solution that satisfies the constraints under the quantum memory capacity $C=35$. The reduction in the entanglement rates as $\tau$ increases maintains the value of $r_j\tau$, $j \in \mathcal{N}$, which keeps the objective function equal to 4.521 for all cases. 
\vspace{-0.3cm}
\subsection{Impact of the Number of Users}
\begin{figure}[t!]
\vskip -0.1in
\begin{center}
\centerline{\includegraphics[scale=0.24]{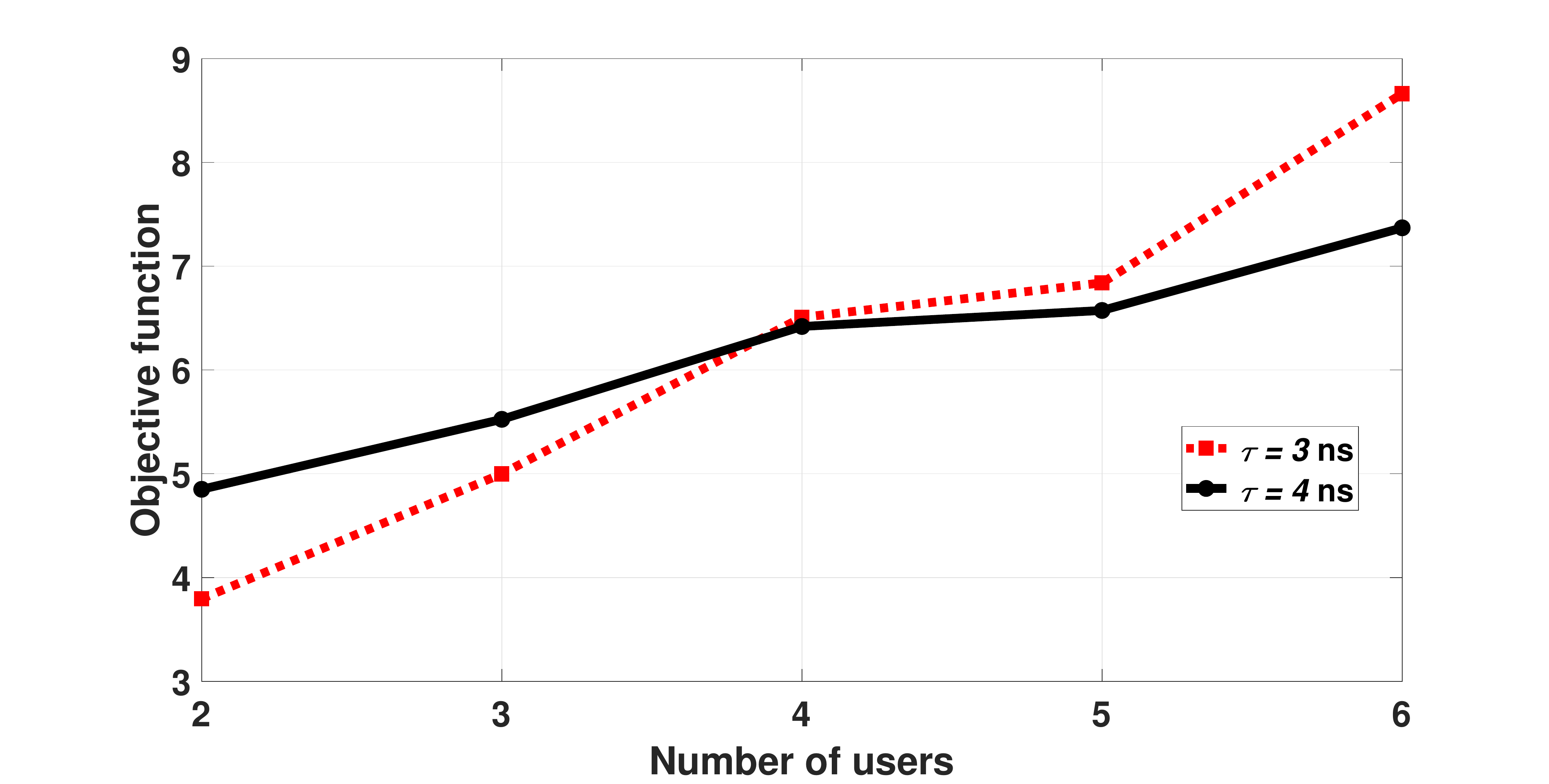}}
\vskip -0.15in
\caption{Number of users vs objective function.}
\label{fig3}
\end{center}
\vskip -0.41in
\end{figure}

In Figure \ref{fig3}, we vary the number of users being served by the quantum node and show how the value of the objective function changes accordingly for different time duration $\tau$. We set $C = 50$, $\epsilon_{\text{min},1} = 1.8\times10^9$ ebit/sec, $\epsilon_{\text{min},2} = 2.2\times10^9$ ebit/sec and for any additional user $j$, $\epsilon_{\text{min},j} = 1.2\times10^9$ ebit/sec. Statistical results are averaged over a large number of runs in which the distances of the users from the quantum node are chosen from a uniform distribution between 0.5 and 5 km. The average achieved objective function values are presented in Figure \ref{fig3}. User 1 is assigned weight $w_1 = 0.8$, while user 2 is assigned weight $w_2 = 0.4$, and any other extra user $j$ is assigned weight $w_j = 0.6$. From Figure \ref{fig3}, we observe that the achieved values of the objective function increase as the number of users increase. Also, when the number of users is small, a larger duration $\tau$ yields a higher value for the objective function. When the number of users becomes greater than or equal to 4, performing the entanglement generation process for a longer duration increases the losses and, thus, decreases the objective function. Another important observation is that the number of users that a quantum node can serve depends on the available quantum memory capacity, the minimum required rate for each user, and the duration of entanglement generation. For example, in some cases for the five-user network when we reduced the memory capacity to 30, there was no feasible solution for the problem.
We also tested for the case in which all users are located at 2 km distance from the quantum node ($d_j = 2$ km $\forall j \in \mathcal{N}$), and are equally weighted $w_j = 1$ $\forall j \in \mathcal{N}$, with $\epsilon_j = 1.2\times10^9$ ebit/sec $\forall j \in \mathcal{N}$, and $\tau = 3$ ns. We observed that the objective function remains the same (equal 4.521) for all numbers of users, however, as we noticed early, the entanglement generation rates vary for each case.
\vspace{-0.2cm}
\subsection{Impact of the Distances}
\begin{figure}[t!]
\begin{center}
\centerline{\includegraphics[scale=0.235]{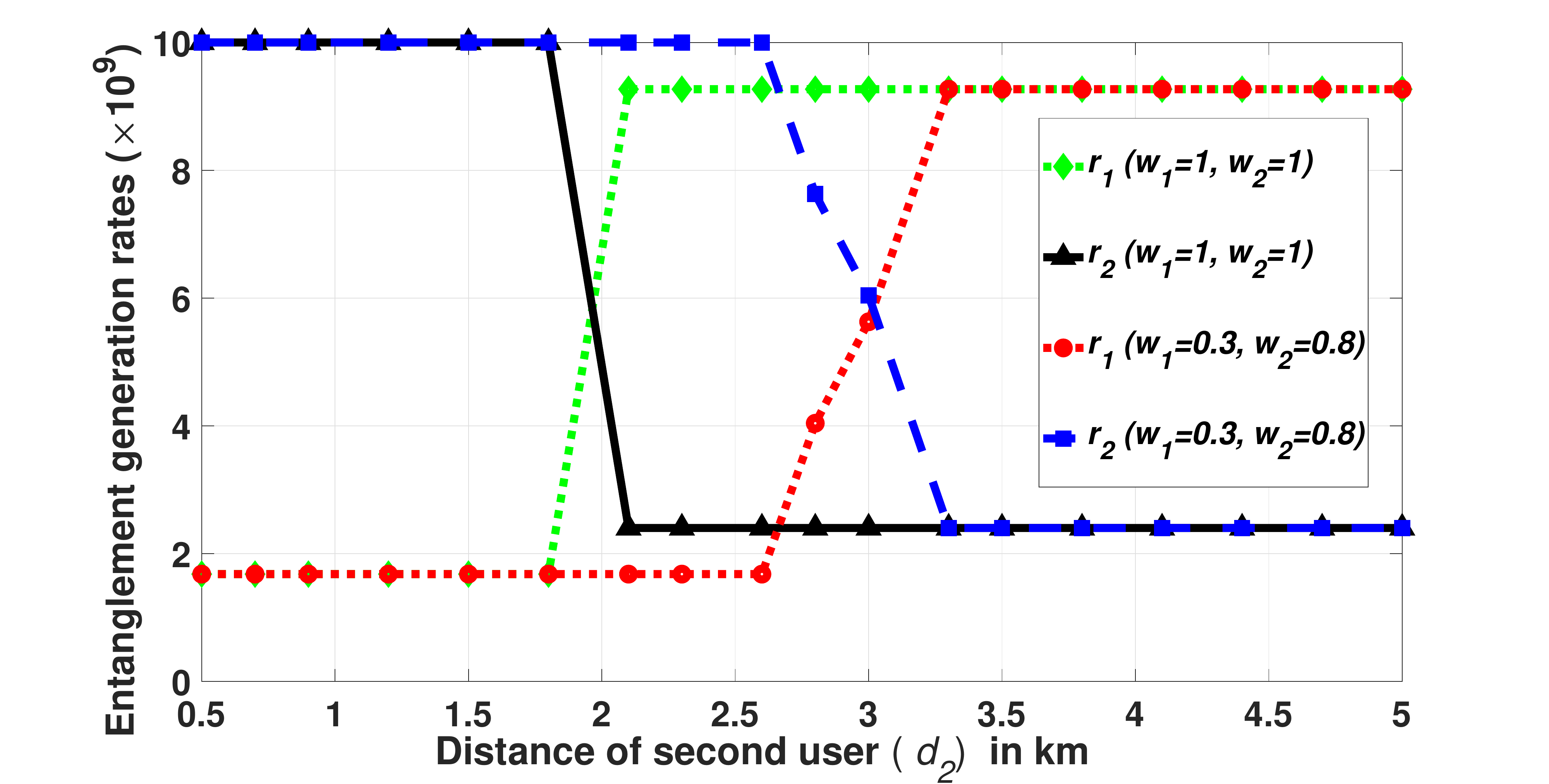}}
\vskip -0.1in
\caption{$d_2$ (km) vs $r_1$ and $r_2$ in ebit/sec for two weighting scenarios.}
\label{fig4}
\end{center}
\vskip -0.4in
\end{figure}

In Figure \ref{fig4}, we consider a network consisting of two users, and we study the effect of varying the distance between one of the users and the quantum node on the entanglement generation rate for each user. We fix $d_1 = 2$ km, and vary $d_2$ for two weighting scenarios. We fix the time duration to $\tau = 3$ ns, and assume the moving user has a larger required minimum value $\epsilon_2 = 2.4\times10^9$ ebit/sec, while $\epsilon_1 = 1.2\times10^9$ ebit/sec. From Figure \ref{fig4}, we can see that the quantum node tends to give the highest possible rate for the user that suffers from a smaller loss, which comes from $w_je^{-\beta_jd_j}$, and the other user is given the minimum required entanglement rate. This figure clearly verifies the importance of the weighting vector in compensating for the loss coming from distance.
\vspace{-0.2cm}
\subsection{Impact of the Quantum Memory Capacity}
\begin{figure}[t!]
\vskip -0.1in
\begin{center}
\centerline{\includegraphics[scale=0.26]{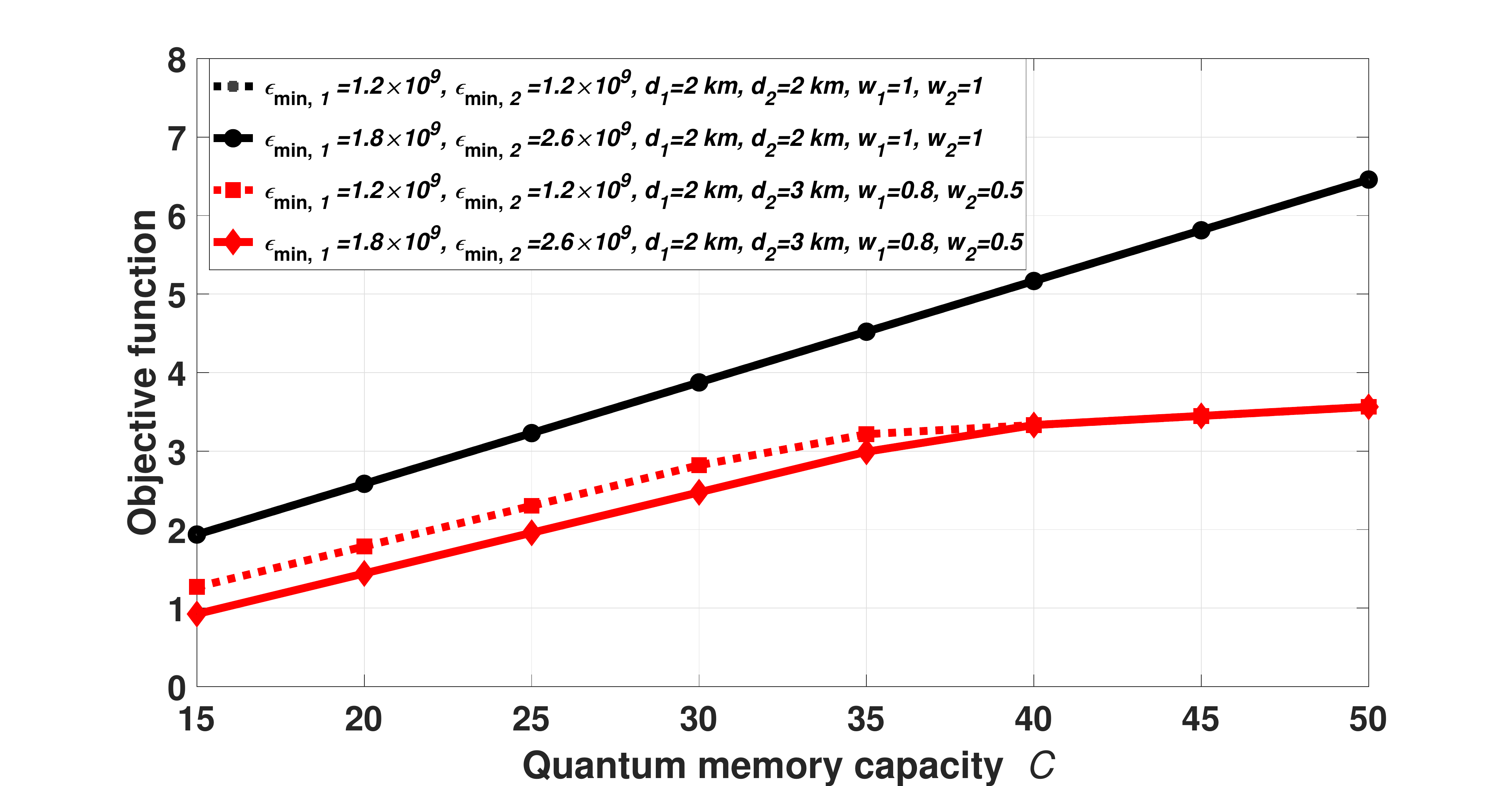}}
\vskip -0.15in
\caption{Quantum Memory Capacity vs Objective Function}
\label{fig5}
\end{center}
\vskip -0.47in
\end{figure}

In Figure \ref{fig5}, we consider a network of two users and vary the quantum memory capacity and notice the effect on the objective function. We consider multiple configurations of the network and notice that as the quantum memory capacity increases, the objective function increases. Also, it is clear that when the two users are equally distant and weighted, changing the minimum entanglement rates $\epsilon_{\text{min},1}$ and $\epsilon_{\text{min},2}$ has no effect on the objective function as $C$ increases. 
\vspace{-0.2cm}
\subsection{Insights}
\label{sec_insights}
Based on the simulation results, we can draw the following insights. First, the number of users that a quantum node can serve heavily relies on the available quantum memory capacity, and, thus, designing memories with higher capacities is a key task. Also, the importance of the weighting we used in our model is verified, since it allows the quantum node to decide on which users to serve more based on their distance from the node and their minimum required entanglement rates. Moreover, the number of users that a quantum node can serve depends on the available quantum memory capacity and the users' requirements, in addition to the entanglement generation period. Thus, in order to effectively design quantum networks, the different users' requirements, their distances from the quantum nodes, and the available memory at each node must be considered in order to decide which users to assign to each node to maximize the profits.
\vspace{-0.15cm}
\section{Conclusion}
\label{sec_conclusion}
\vspace{-0.1cm}
In this paper, we have studied the problem of optimal distribution of entanglement generation for multiple heterogeneous users in a quantum communication network. This is an important problem for the next generation 6G wireless communication and represents a step towards achieving the envisioned quantum Internet. We have formulated a mixed integer nonlinear programming optimization problem where the entanglement generation rates, which are a key resource in quantum communication, are optimized in order to efficiently utilize the available quantum memory capacity and to achieve fairness among users. We have solved the problem using interior point optimization solvers. Simulation results have shown the effectiveness of the proposed system model in designing efficient quantum networks.
\vspace{-0.15cm}

\bibliographystyle{IEEEtran}
\bibliography{References}

\end{document}